\documentclass{article}

\usepackage{arxiv}

\usepackage{times}
\usepackage{epsfig}
\usepackage{graphicx}
\usepackage{amsmath}
\usepackage{amssymb}

\usepackage{adjustbox}
\newcommand{\widthscalefive}{0.10}

\usepackage{pifont}% http://ctan.org/pkg/pifont
\usepackage[utf8]{inputenc} % allow utf-8 input
\usepackage[T1]{fontenc}    % use 8-bit T1 fonts
\usepackage{hyperref}       % hyperlinks
\usepackage{url}            % simple URL typesetting
\usepackage{booktabs}       % professional-quality tables
\usepackage{amsfonts}       % blackboard math symbols
\usepackage{nicefrac}       % compact symbols for 1/2, etc.
\usepackage{microtype}      % microtypography
\usepackage{cleveref}       % smart cross-referencing
\usepackage{lipsum}         % Can be removed after putting your text content
\usepackage{graphicx}
\usepackage{natbib}
\usepackage{doi}

\title{Adaptive Single Image Deblurring}

\date{}

\author{ \hspace{1mm} Maitreya Suin
	\And
	\hspace{1mm} Kuldeep Purohit
	\And
	\hspace{1mm} A. N. Rajagopalan
}

\renewcommand{\headeright}{}
\renewcommand{\undertitle}{}
\begin{document}
\maketitle

\begin{abstract}
	This paper tackles the problem of dynamic scene deblurring. Although end-to-end fully convolutional designs have recently advanced the state-of-the-art in non-uniform motion deblurring, their performance-complexity trade-off is still sub-optimal. Existing approaches achieve a large receptive field by a simple increment in the number of generic convolution layers, kernel-size, which comes with the burden of the increase in model size and inference speed. In this work, we propose an efficient pixel adaptive and feature attentive design for handling large blur variations within and across different images. We also propose an effective content-aware global-local filtering module that significantly improves the performance by considering not only the global dependencies of the pixel but also dynamically using the neighboring pixels. We use a patch hierarchical attentive architecture composed of the above module that implicitly discover the spatial variations in the blur present in the input image and in turn perform local and global modulation of intermediate features. Extensive qualitative and quantitative comparisons with prior art on deblurring benchmarks demonstrate the superiority of the proposed network. 
\end{abstract}

% keywords can be removed

\section{Introduction}
Blind motion deblurring is an ill-posed problem which aims to recover a sharp image from a given image degraded due to motion based smearing of texture and high-frequency details. Due to its diverse applications in surveillance, remote sensing, and cameras mounted on hand-held and vehicle mounted cameras, deblurring has gathered substantial attention from computer vision and image processing community in past two decades. 

Majority of existing deblurring approaches are based on variational model, whose key component is the regularization term. Large literature studies design of priors that are apt for recovering the underlying undistorted image and the camera trajectory. The restoration quality depends on the selection of the prior, its weight, as well as tuning of other parameters involving highly non-convex optimization setups (\cite{nimisha2017blur}). A significant number of works have been proposed \cite{paramanand2011depth,nimisha2018unsupervised,rao2014harnessing,nimisha2018generating,vasu2017local,paramanand2014shape,vijay2013non} where various traditional approaches were adopted for deblurring. Non-uniform blind deblurring for general dynamic scenes is a challenging computer vision problem as blurs arise from various sources including moving objects, camera shake and depth variations, causing different pixels to capture different motion trajectories. Such hand-crafted priors struggle while generalizing across different types of real-world examples, where blur is far more complex than modeled \cite{gong2017motion}. 

% Regarding motion trajectory, various conventional and learning based approaches exist [18][16][12] for improving the blur kernel estimation step, but they assume only translational camera motion, which significantly restricts their utility. Several extensions consider rotations $r_x$,$r_y$ and $r_z$ to parametrize the 3D motion of the camera. Traditional deblurring approaches made use of parametric blur models to estimate kernels. An exhaustive survey of blind deblurring algorithms can be found in \cite{lai2016comparative}.

Non-uniform blind deblurring for general dynamic scenes is a challenging computer vision problem as blurs arise from various sources including moving objects, camera shake and depth variation, causing different pixels to capture different motion trajectories. Conventional hand-designed formation models would require explicit estimation of all of these independent variables from a single blurred image, which is an extremely ill-posed problem. As a result, applying such algorithms on general dynamic scenes yields images with unpleasant artifacts and incomplete deblurring. 

\par Recent works \cite{purohit2019bringing,purohit2020region,mohan2021deep,mohan2019unconstrained,vasu2018non} based on deep convolutional neural networks (CNN) have studied the benefits of replacing the image formation model with a parametric model that can be trained to emulate the non-linear relationship between blurred-sharp image pairs. Such works~\cite{nah2017deep} directly regress to deblurred image intensities and overcome the limited representative capability of variational methods in describing dynamic scenes. These methods can handle combined effects of camera motion and dynamic object motion and achieve state-of-the-art results on single image deblurring task. They have reached a respectable reduction in model size, but still lack in accuracy and are not real-time.

Reaching a trade-off between the size and speed of the network, the receptive field and the accuracy of restoration is a non-trivial task. Therefore, our work focuses on the design of efficient modules that achieve performance improvement without requiring a deeper framework. We investigate motion-dependent adaptability within a CNN. This problem is challenging especially since multiple segments of different sizes and motion co-exist in a single image. Since motion blur is essentially a direct aggregation of spatial transformation of the image, a deblurring network can benefit from adapting to the magnitude as well as the direction of motion. We deploy content-aware modules which adjusts the filter to be applied and the context of each pixel depending on the motion information.

Following the state of the art in deblurring, we adopt an multi-patch hierarchical design to directly estimate the restored sharp image. Instead of cascading along depth, we introduce content-aware feature and filter transformation capability through global-local attentive module and residual attention across layers to improve their performance. These modules learn to exploit the similarity in the motion between different pixels within an image and also sensitive to position specific local context. 

Our design originates from the intuition that motion blur is essentially a aggregation of various spatially varying transformation of the sharp image, and hence a deblurring network can benefit from implicitly decoding the magnitude as well as the direction of motion accompanied with the global context. The proposed module can dynamically generate global attention map and local filters almost in real-time. The feature transformations and filters estimated by the network are image dependent and hence can be visualized for different images. The efficiency of our architecture is demonstrated through comprehensive comparison on two benchmarks with the state-of-the-art deblurring approaches. Our model achieves superior performance, while being computationally more efficient, resulting in real-time deblurring of HD images on a single GPU. The major contributions in this work are:
\begin{enumerate}
	\item We propose an efficient deblurring design built on new convolutional modules that learn transformation of features using global attention and dynamic local filters. We show that these two branches complements each other and results in superior deblurring performance. Moreover, the efficient design of attention-module enables us to use it throughout the network without the need of explicit downsampling.

	\item We provide extensive analysis and evaluations on static and dynamic scene deblurring benchmarks.
\end{enumerate}

\section{Method}
\begin{figure}[t]
	\begin{center}
		\includegraphics[width=0.7\linewidth,height = 0.5\linewidth]{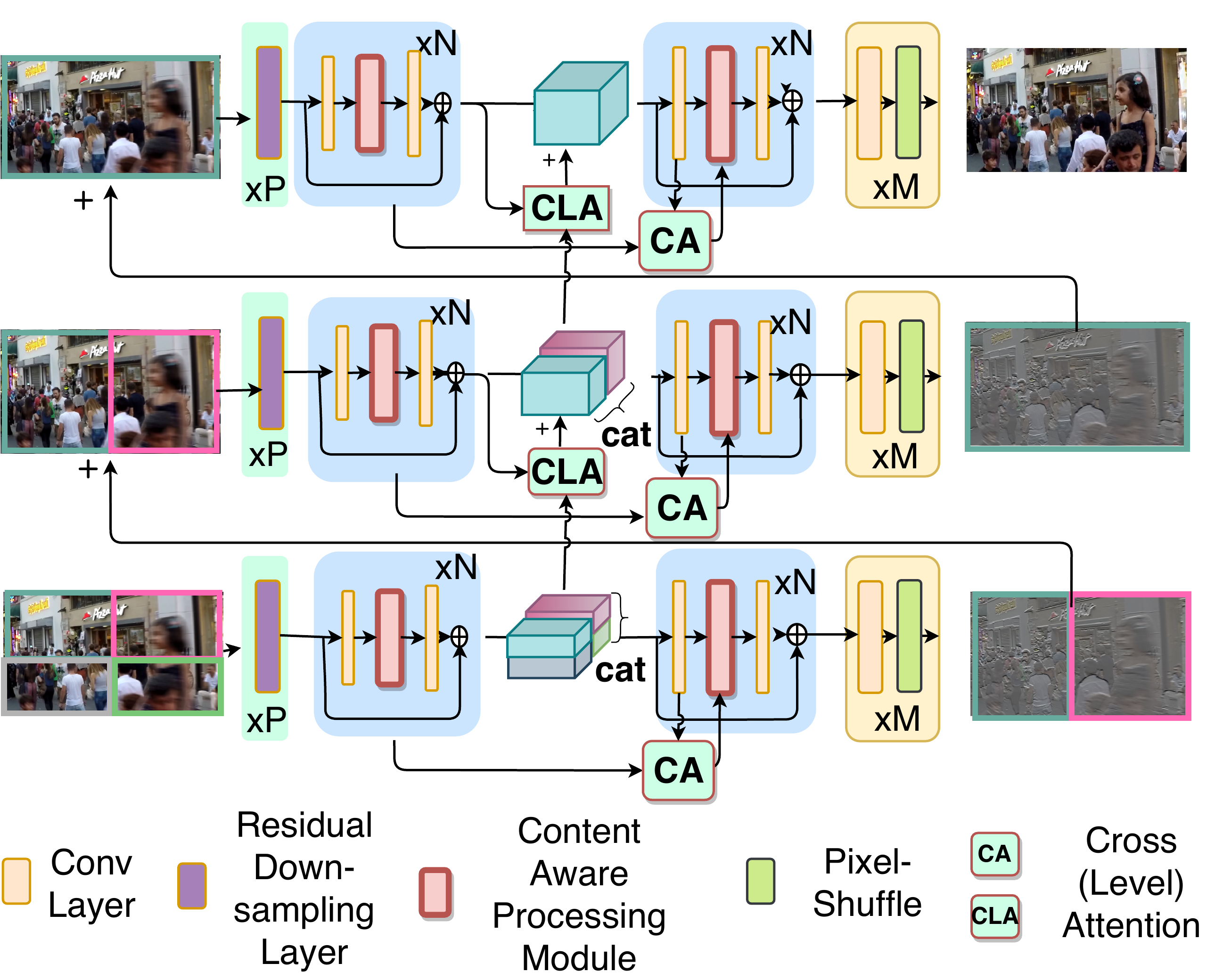}
	\end{center}
	\caption{Overall architecture of our proposed network.}
	\label{fig:mainarch}
\end{figure}

To date, the driving force behind performance improvement in deblurring has been the use of a large number of layers and larger filters which assist in increasing the "static" receptive field and the generalization capability of a CNN. However, these techniques offer suboptimal design, since network performance does not always scale with network depth, as the effective receptive field of deep CNNs is much smaller than the theoretical value (investigated in \cite{luo2016understanding}).

Although previous multi-scale and scale-recurrent methods have shown good performance in removing non-uniform blur, they suffer from expensive inference time and performance bottleneck while simply increasing model depth. Instead, inspired by \cite{zhang2019deep} , we adopt multi-patch hierarchical structure as our base-model, which compared to multi-scale approach has the added advantage of residual-like architecture that leads to efficient learning and faster processing speed.

The overall architecture of our proposed network is shown in Fig. \ref{fig:mainarch}. We divide the network into 3 levels instead of 4 as described in \cite{zhang2019deep}. We found that the relative performance gain due to the inclusion of level 4 is negligible compared to the increase in inference time and number of parameters. At the bottom level input sliced into 4 non-overlapping patches for processing, and as we gradually move towards higher levels, the number of patches decrease and lower level features are adaptively fused using attention module as shown in Fig. \ref{fig:mainarch}. The output of level 1 is the final deblurred image. Note that unlike \cite{zhang2019deep}, we also avoid cascading of our network along depth, as that adds severe computational burden. Instead, we advocate the use of content-aware processing modules which yield significant performance improvements over even the deepest stacked versions of original DMPHN \cite{zhang2019deep}. Major changes incorporated in our design are described next.

\par Each level of our network consists of an encoder and a decoder. Both the encoder and the decoder are made of standard convolutional layer and residual blocks where each of these residual blocks contains 1 convolution layer followed by a content-aware processing module and another convolutional layer. The content-aware processing module comprises two branches for global and local level feature processing which are dynamically fused at the end. The residual blocks of decoder and encoder are identical except for the use of cross attention in decoder. We have also designed cross-level attention for effective propagation of lower level features throughout the network. We begin with describing content-aware processing module, then proceed towards the detailed description of the two branches and finally how these branches are adaptively fused at the end.

\subsection{Attention}
Given the input $x_m$, we generate three attention maps $P \in \mathbb{R}^{C_2 \times HW}$ , $Q \in \mathbb{R}^{C_2 \times HW}$ and $M_2 \in \mathbb{R}^{C}$ using convolutional operations $f_p(\cdot)$ ,$f_q(\cdot)$ and $f_{M_2}(\cdot)$ where global-average-pooling is used for the last case to get $C$ dimensional representation.  We take the first cluster of attention map $Q$ and split it into $C_2$ different maps $Q =\{q_1,q_2,...,q_{C_2}\}$, $q_i \in \mathbb{R}^{HW}$ and these represent $C_2$ different spatial attention-weights. A single attention reflects one aspect of the blurred image. However, there are multiple pertinent properties like edges,textures etc. in the image  that together helps removing the blur. Therefore, we deploy a cluster of attention maps to effectively gather $C_2$ different key features. Each attention map is element-wise multiplied with the input feature map $x_{m_1}$ to generate $C_2$ part feature maps as
\begin{equation}
	x^k_{m_1} = q_k \odot x_{m_1} ~ ~ ~ , \text{with}  \sum_{i=1}^{HW} q_{ki} = 1 ~ ~ ~ ~ ~ ~(k= 1,2,...,N)
\end{equation}
where $x^k_m \in \mathbb{R}^{C \times HW}$. We further extract descriptive global feature by global-sum-pooling (GSP) along $HW$ dimension to obtain $k^{th}$ feature representation as 
\begin{equation}
	\bar{x}^k_{m_1} = GSP_{HW}(x^k_{m_1}) ~ ~ ~ ~ ~ ~(k= 1,2,...,N)
\end{equation}
where $\bar{x}^k_m \in \mathbb{R}^C$. Now we have $\bar{x}_{m_1} = \{\bar{x}^1_{m_1},\bar{x}^2_{m_1},...,\bar{x}^{C_2}_{m_1}\}$ which are obtained from $C_2$ different attention-weighted average of the input $x_m$. Each of these $C_2$ representations is expressed by an $C$-dimensional vector which is a feature descriptor for the $C$ channels. We further enhance these $C$ dimensional vectors by emphasizing the important feature-embeddings as
\begin{equation}\label{eq:sa2}
	\bar{x}^k_{{m_1}{m_2}} = M_2 \odot \bar{x}^k_{m_1}
\end{equation}
where $M_2$ can be expressed as
\begin{equation}
	M_2 = f_{m_2}(\bar{x}_{m_1};\theta_{m_2}) \in \mathbb{R}^C
\end{equation}
Next we take the set of attention maps $P = \{p_1,p_2,...,p_{HW}\}$ where $p_i \in \mathbb{R}^{C_2}$ is represents attention map for $i^{th}$ pixel. Intuitively, $p_i$ shows the relative importance of $C_2$ different attention-weighted average ($\bar{x}_{{m_1}{m_2}}$) for the current pixel and it allows the pixel to adaptively select the weighted average of all the pixels. For each output pixel $j$, we element-wise multiply these $C_2$ feature representations $\bar{x}^k_{{m_1}{m_2}}$ with the corresponding attention map $p_j$, to get
\begin{equation}
	y^j = p_j \odot \bar{x}_{{m_1}{m_2}} ~ ~ \text{with}  \sum_{i=1}^{C_2} p_{ji} = 1 ~ ~ , (j= 1,2,...,HW)
\end{equation}
where ${y}^j \in \mathbb{R}^{C \times C_2}$. We again apply global-average-pooling on ${y}^j$ along $C_2$ to get $C$ dimensional feature representation for each pixel as
\begin{equation}
	\bar{y}^{j} = GAP_{C_2}({y}^j)
\end{equation}
where $\bar{y}^{j} \in \mathbb{R}^C$ represent the accumulated global feature for the $j^{th}$ pixel. 

\begin{figure*}[htb] \label{fig:visual_deblur}
	\scriptsize
	\centering
	%\tiny
	\begin{tabular}{cccccccc}
		\includegraphics[width=0.10\textwidth]{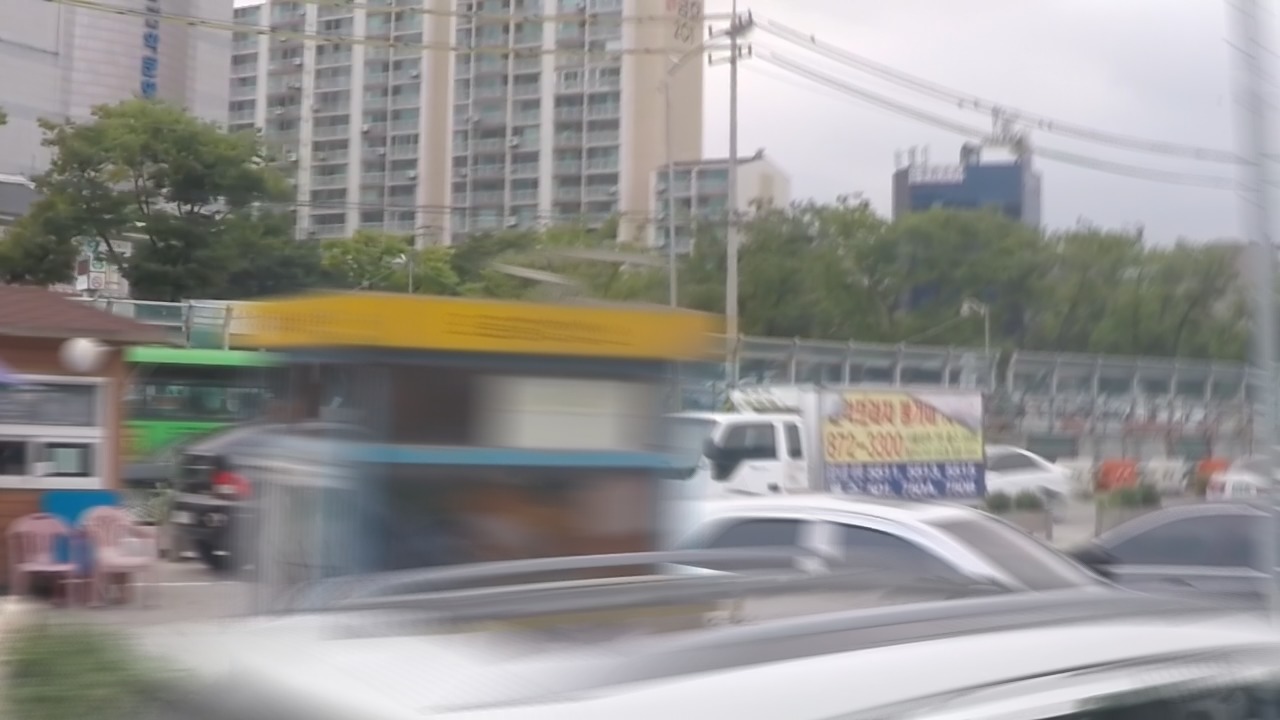} &
		\includegraphics[width=\widthscalefive \textwidth]{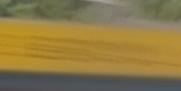} & 
		\includegraphics[width=\widthscalefive \textwidth]{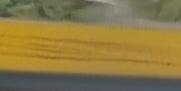} & %\hspace{\fsdttwofig} &
		\includegraphics[width=\widthscalefive \textwidth]{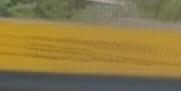} & %\hspace{\fsdttwofig} &
		\includegraphics[width=\widthscalefive \textwidth]{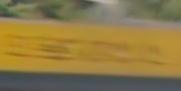} &
		\includegraphics[bb=450 360 620 450,clip=True,width=\widthscalefive \textwidth]{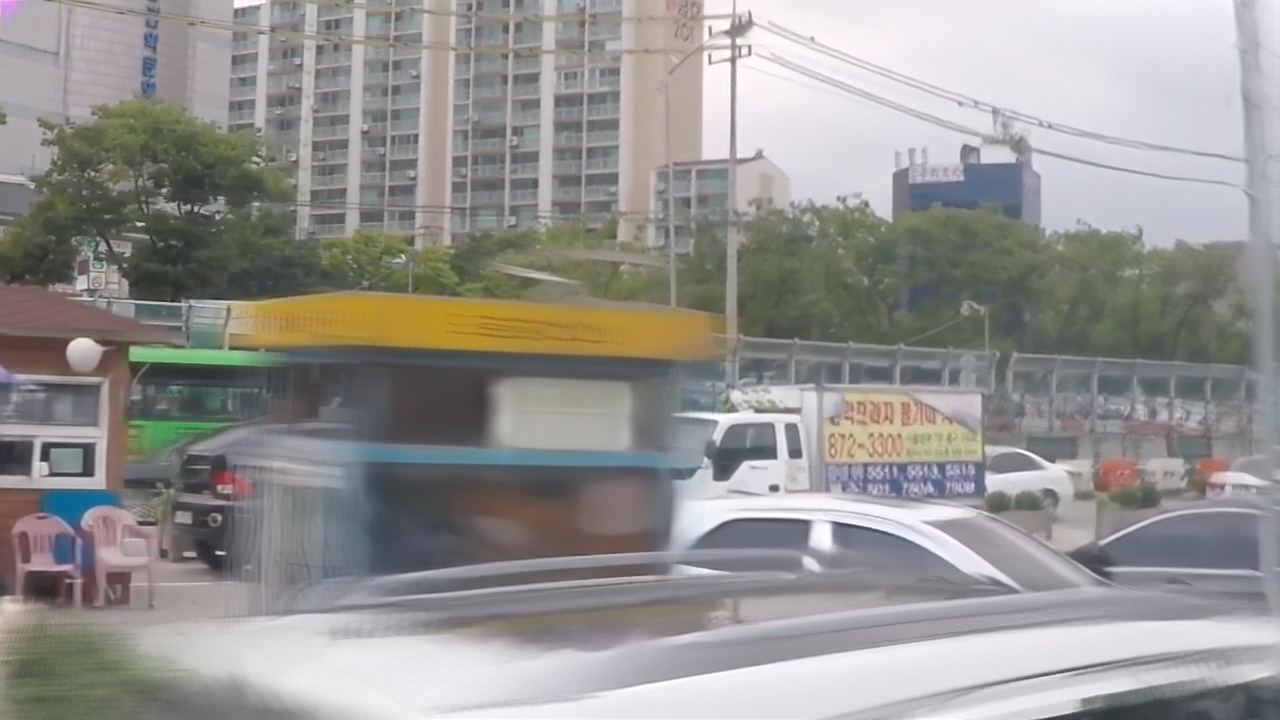} &
		\includegraphics[width=\widthscalefive \textwidth]{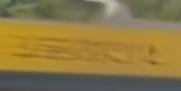} & %\hspace{\fsdttwofig} &
		\includegraphics[width=\widthscalefive \textwidth]{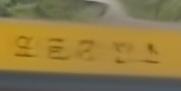}
		\\
		\includegraphics[width=0.10\textwidth]{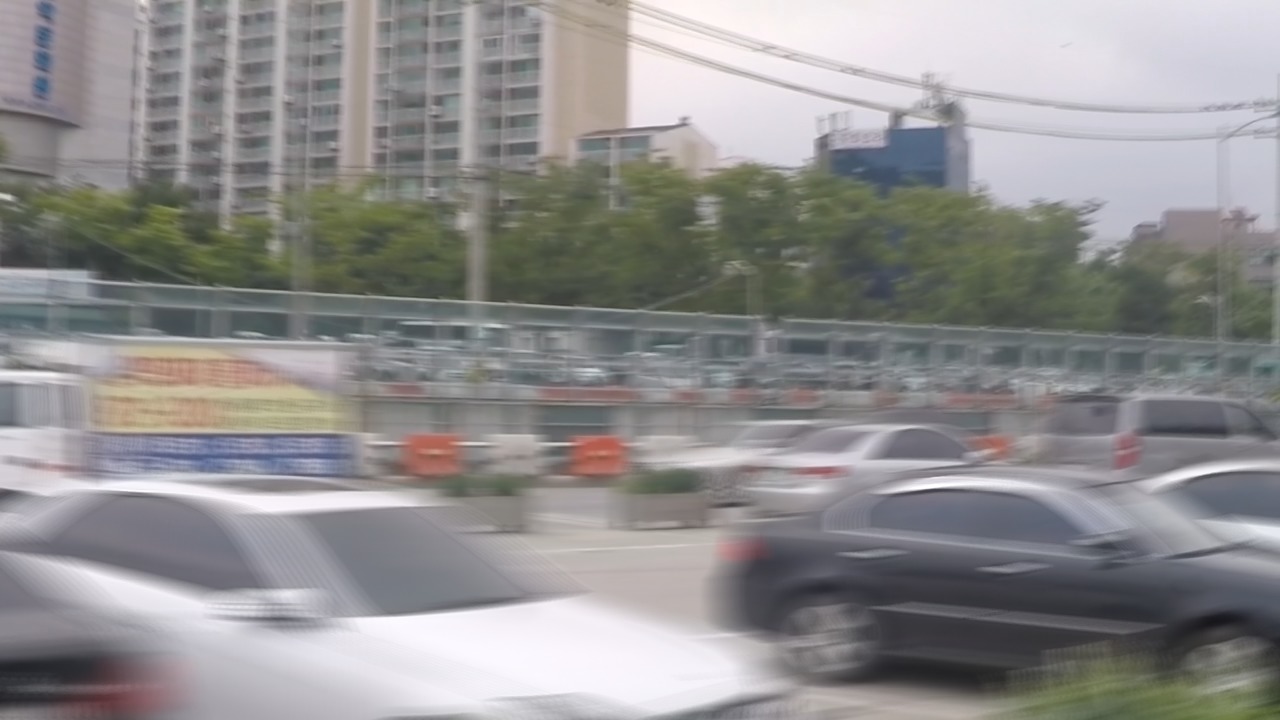}
		&
		\includegraphics[width=\widthscalefive \textwidth]{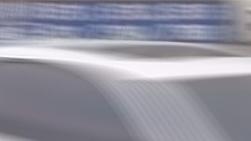} & %\hspace{\fsdttwofig} &
		
		\includegraphics[width=\widthscalefive \textwidth]{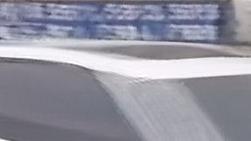} & %\hspace{\fsdttwofig} &
		\includegraphics[width=\widthscalefive \textwidth]{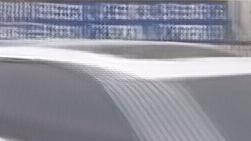} & %\hspace{\fsdttwofig} &
		\includegraphics[width=\widthscalefive \textwidth]{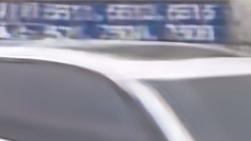} &
		\includegraphics[bb=80 90 390 280,clip=True,width=\widthscalefive \textwidth]{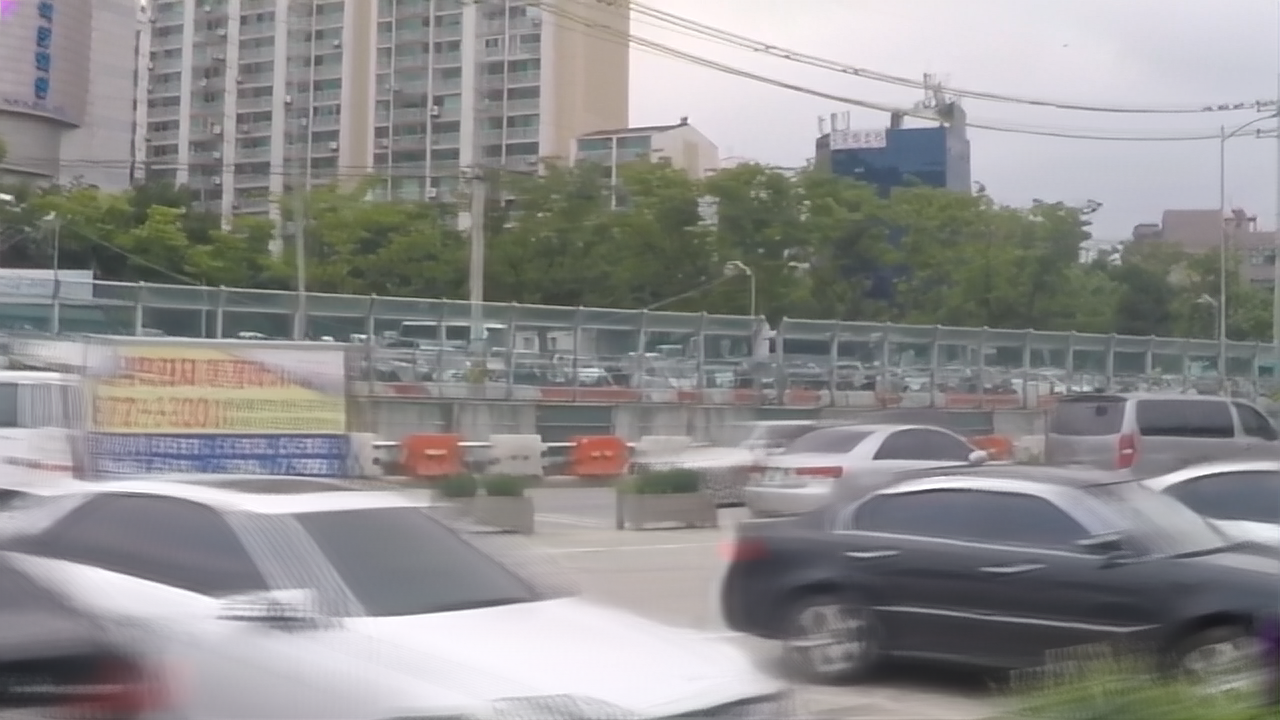} &
		\includegraphics[width=\widthscalefive \textwidth]{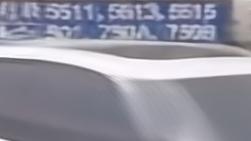} &				 %\hspace{\fsdttwofig} &3
		\includegraphics[width=\widthscalefive \textwidth]{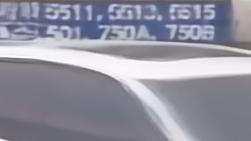}				%\vspace{-3mm}
		\\
		\\
		\includegraphics[width=0.10\textwidth]{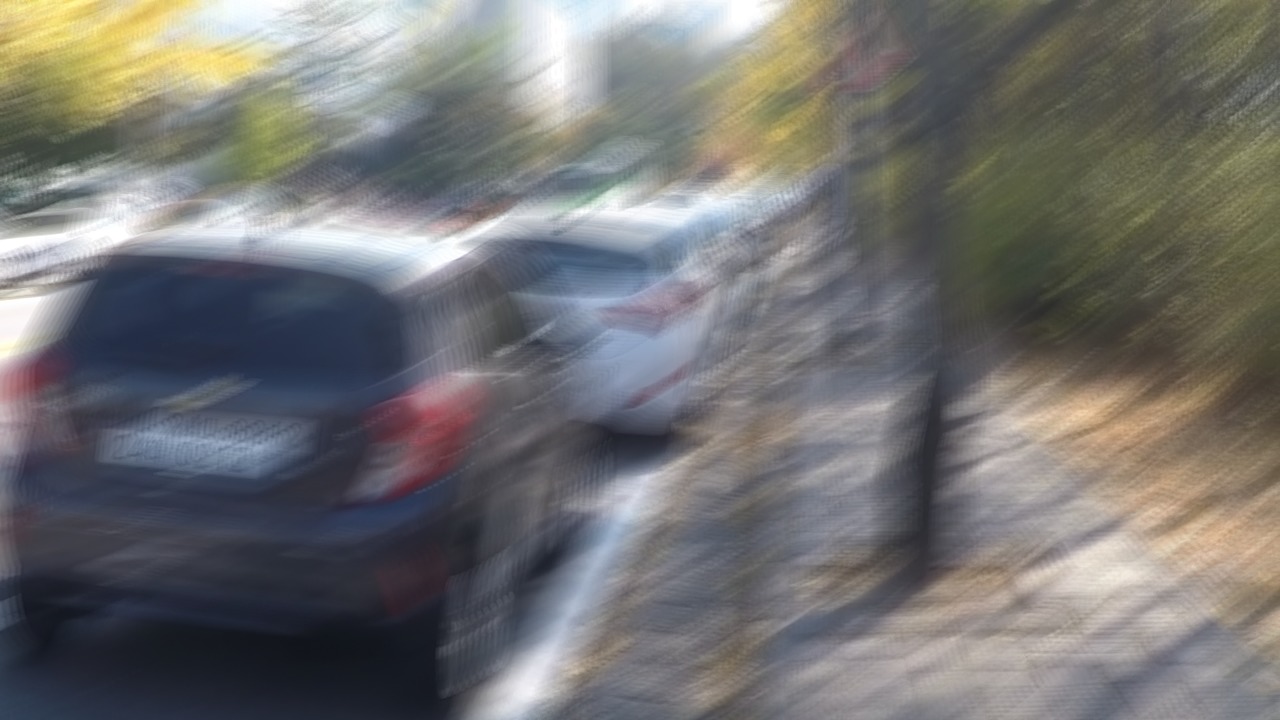}
		&
		\includegraphics[width=\widthscalefive \textwidth]{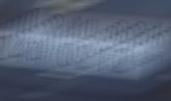} & %\hspace{\fsdttwofig} &
		
		\includegraphics[width=\widthscalefive \textwidth]{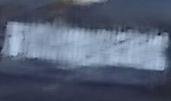} & %\hspace{\fsdttwofig} &
		\includegraphics[width=\widthscalefive \textwidth]{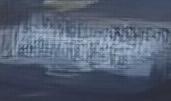} & %\hspace{\fsdttwofig} &
		\includegraphics[width=\widthscalefive \textwidth]{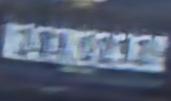} &
		\includegraphics[bb=120 220 320 330,clip=True,width=\widthscalefive \textwidth]{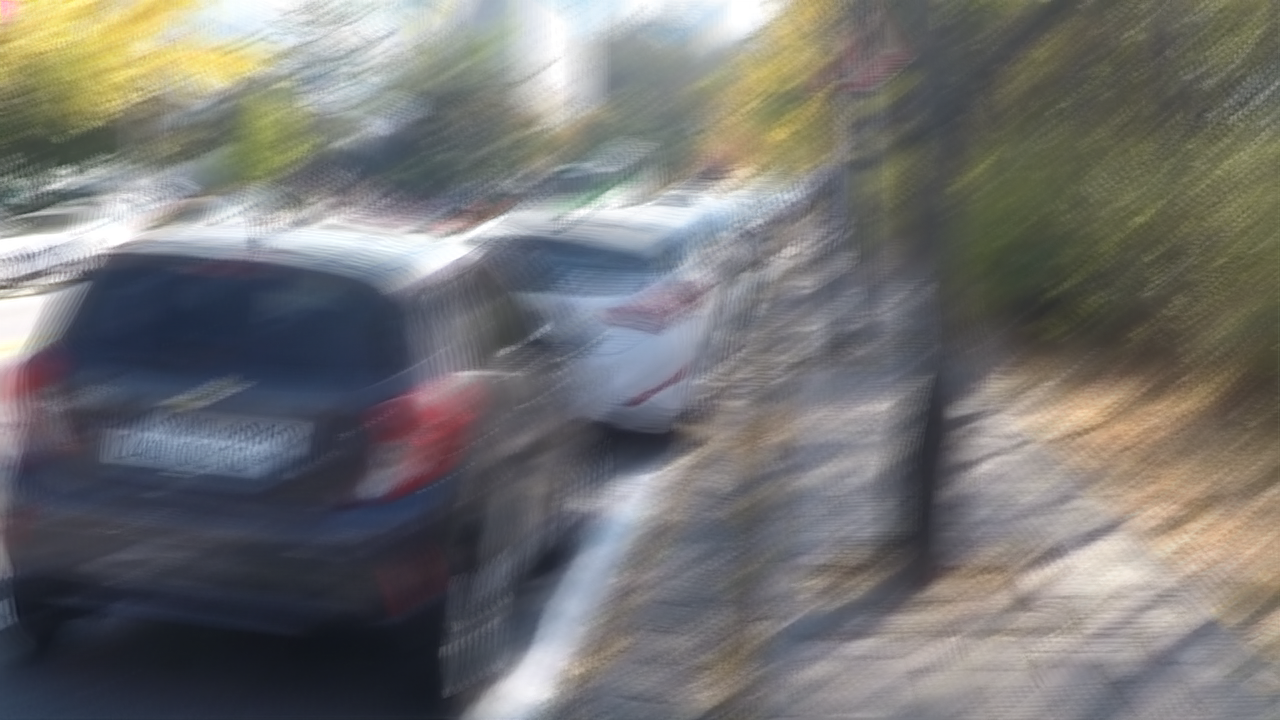} &
		\includegraphics[width=\widthscalefive \textwidth]{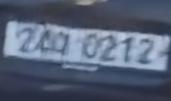} & 				%\hspace{\fsdttwofig} &3
		\includegraphics[width=\widthscalefive \textwidth]{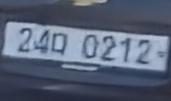}				%\vspace{-3mm}				
		
		\\ 
		\includegraphics[width=0.10\textwidth]{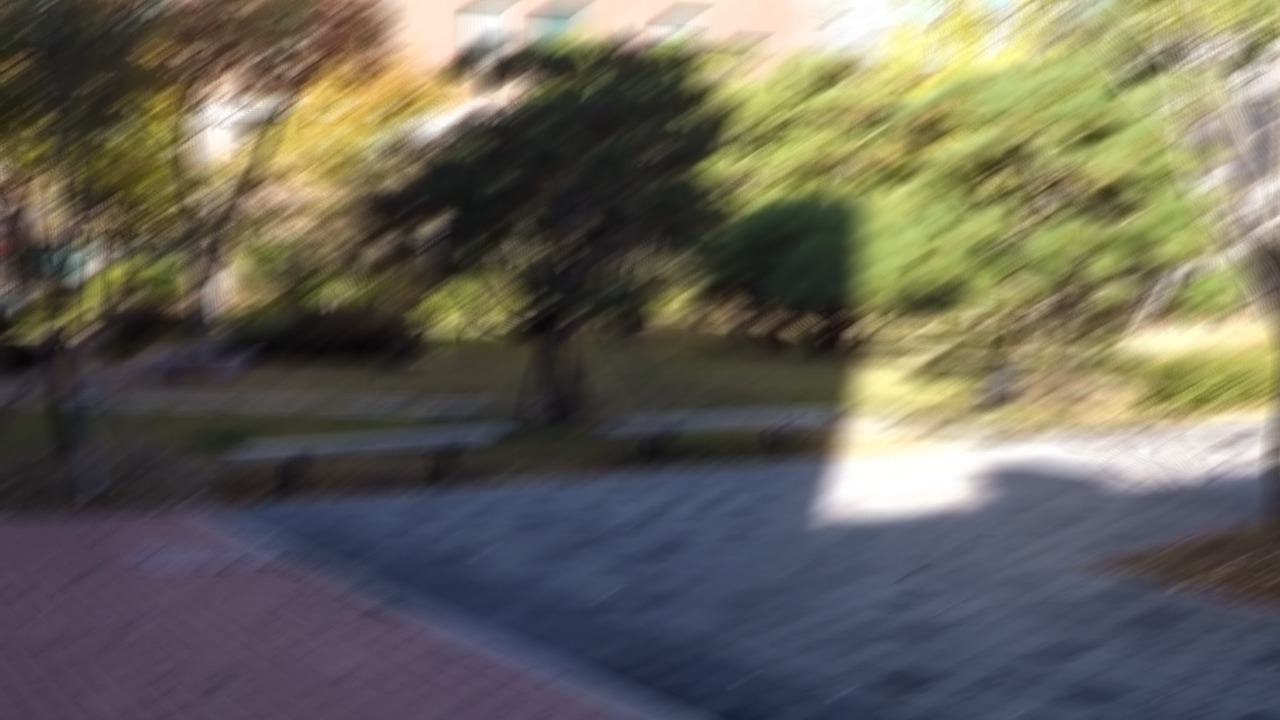}
		&
		\includegraphics[width=\widthscalefive \textwidth]{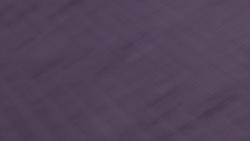} & %\hspace{\fsdttwofig} &
		
		\includegraphics[width=\widthscalefive \textwidth]{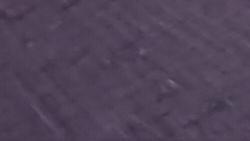} & %\hspace{\fsdttwofig} &
		\includegraphics[width=\widthscalefive \textwidth]{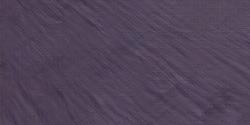} & %\hspace{\fsdttwofig} &
		\includegraphics[width=\widthscalefive \textwidth]{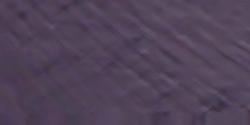} &
		\includegraphics[bb=10 0 300 130,clip=True,width=\widthscalefive \textwidth]{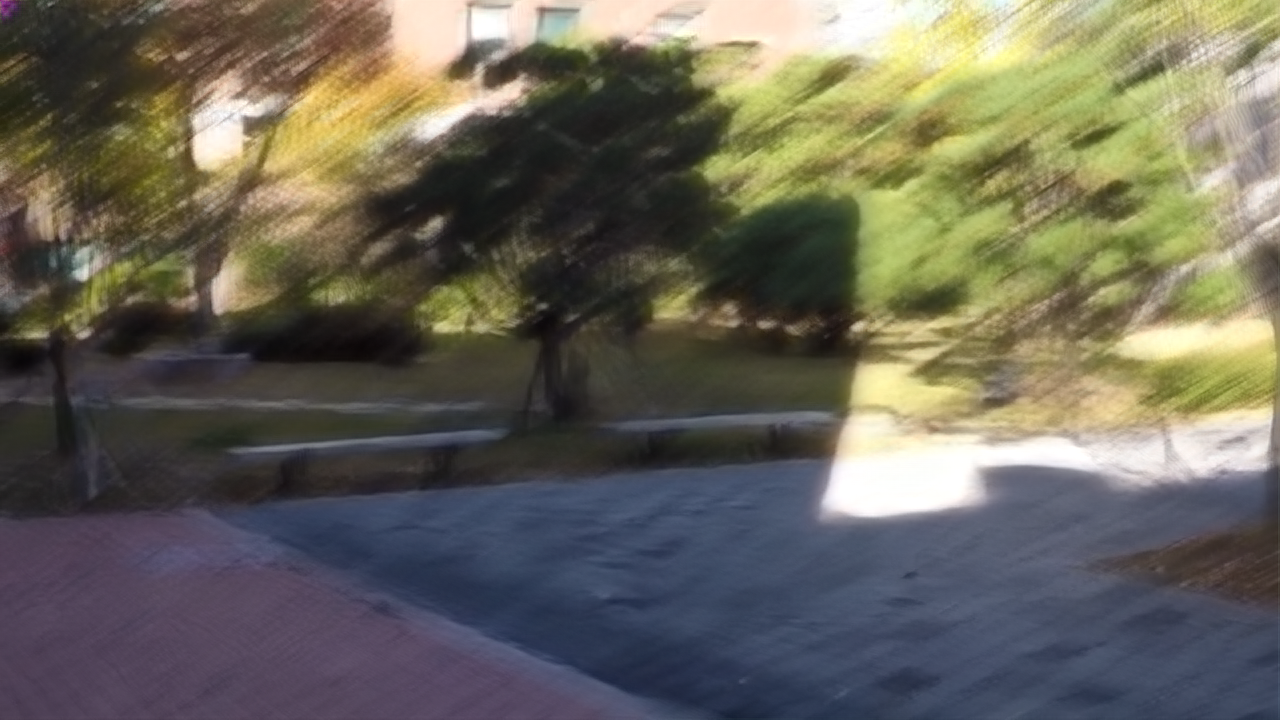} &
		\includegraphics[width=\widthscalefive \textwidth]{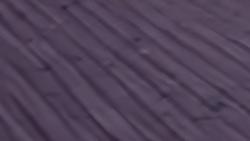} & 				%\hspace{\fsdttwofig} &3
		\includegraphics[width=\widthscalefive \textwidth]{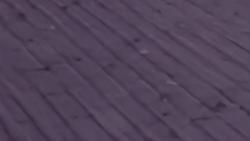}				%\vspace{-3mm}				
		
		\\ 				
		(a) Blurred Image&
		
		(b) Blurred patch& %\hspace{\fsdttwofig} &
		(c) MS-CNN & %\hspace{\fsdttwofig} &
		(d) DelurGAN& % \hspace{\fsdttwofig} &
		(e) SRN & %\hspace{\fsdttwofig} &
		(f) DelurGAN-V2 & % \hspace{\fsdttwofig} &
		(g) \tiny{Stack(4)-DMPHN} & %\hspace{\fsdttwofig} &
		(h) Ours%\hspace{\fsdttwofig} &
		\\
	\end{tabular}
	%\vspace{-0em}
	\caption{Visual comparisons of deblurring results on images from the GoPro test set~\cite{nah2017deep}. Key blurred patches are shown in (b), while zoomed-in patches from the deblurred results are shown in (c)-(h).}% (best viewed in high resolution).}
\label{fig:dynamic}
%\vspace{-0em}
\end{figure*} 

\subsection{Filtering}
We design a content-aware "global-local" processing module which depending on the input, deploys 2 parallel branch to fuse global and local features. The global branch is made of the the attention module described above. For decoder this includes both self and cross module attention whereas for encoder only self-attention is used. In contrary to \cite{bello2019attention}, for the local branch we use a content-aware convolutional layer where the filter is spatially varying and changes at runtime depending on the input image to handle spatially-varying dynamic motion blur effectively. Our work is based on \cite{su2019pixel} as we use a ‘meta-layer’ to generate pixel dependent spatially varying kernel to implement spatially variant convolution operation. Given the input feature map $x \in  \mathbb{R}^{C \times H \times W}$, we apply an kernel generation funtcion to generate a spatially varying kernel $K$ and do the convolution operation as:
\begin{equation}
	x^{dyn}_i  = \sum_{j \in \Omega(i)} K_{p_i,p_j} W[p_i - p_j] x_j + b ~ ~ ~ ~ (i= 1,2,...,HW)
\end{equation}
where $x^{dyn}_i \in \mathbb{R}^C$, $p_i= (x_i,y_i)^T$ are pixel coordiantes, $\Omega()$ defines the convolution window, $b$ denotes biases and $K_{p_i,p_j}$ is the pixel dependent kernel generated. Standard spatial convolution can be seen as a special case of the above with adapting kernel being constant $K_{p_i,p_j} = 1$.
Finally we sum-fuse the output of these two branches as
\begin{equation}
	x^{GL} = x^{att} + x^{dyn}
\end{equation}

\section{Experiments}
\subsection{Implementation Details}
\noindent \textbf{Datasets:} We follow the configuration of~\cite{zhang2019deep,kupyn2019deblurgan,tao2018scale,kupyn2017deblurgan,nah2017deep}, which train on 2103 images from the GoPro dataset (\cite{nah2017deep}).

\noindent\textbf{Training settings and implementation details:} 
All the convolutional layers within our proposed modules contain $128$ filters. The hyper-parameters for our encoder-decoder backbone are $N=3$, $M=2$, and $P=2$, and filter size in PDF modules is $5\times5$. Following \cite{zhang2019deep}, we use batch-size of $6$ and patch-size of $256\times256$. Adam optimizer~\cite{kingma2014adam} was used with initial leaning rate $10^{-4}$, halved after every $2\times10^{5}$ iterations. We use PyTorch 
(\cite{paszke2017automatic}) library and Titan Xp GPU.

\begin{table*}[htbp] \label{tbl:gopro}
	\centering
	\caption{Performance comparisons with existing algorithms on 1103 images from the deblurring benchmark GoPro \cite{nah2017deep}.\label{TableGopro}}
	\resizebox{0.4\textwidth}{!}{
		\begin{tabular}{|c|c|c|c|}
			\hline
			Method & PSNR & SSIM & Time \\
			\hline 
			\cite{xu2013unnatural} & 21 & 0.741 & 3800 \\
			\cite{whyte2012non} & 24.6 & 0.846 & 700\\
			\cite{hyun2013dynamic} & 23.64 & 0.824 & 3600\\
			\cite{gong2017motion} &  26.4 & 0.863 & 1200\\
			\cite{nah2017deep} & 29.08 & 0.914 & 6\\
			\cite{kupyn2017deblurgan} & 28.7 & 0.858 & 1\\
			\cite{tao2018scale} & 30.26 & 0.934 & 1.2\\
			\cite{zhang2018dynamic}  & 29.19 & 0.931 &1\\
			\cite{gao2019dynamic} &30.90 &0.935 &1.0\\
			\cite{zhang2019deep} & 31.20 &0.940 &0.98\\
			\cite{kupyn2019deblurgan} &29.55  &0.934 & 0.48\\
			Ours & 31.85 & 0.948 & 0.34\\
			\hline

		\end{tabular}
	}
\end{table*}

\subsection{Performance comparisons}
The main application of our work is efficient deblurring of general dynamic scenes. Due to the complexity of the blur present in such images, conventional image formation model based deblurring approaches struggle to perform well. Hence, we compare with only two conventional methods \cite{whyte2012non,xu2013unnatural} (which are selected as representative traditional methods for non-uniform deblurring, with publicly available implementations). We provide extensive comparisons with state-of-the-art learning-based methods, namely MS-CNN\cite{nah2017deep}, DeblurGAN\cite{kupyn2017deblurgan}, DeblurGAN-v2\cite{kupyn2019deblurgan}, SRN\cite{tao2018scale}, and Stack(4)-DMPHN\cite{zhang2019deep}. We use official implementation from the authors with default parameters.

\textbf{Quantitative Evaluation}
We show performance comparisons on two different benchmark datasets. The quantitative results on GoPro testing set are listed in Table 1.

The average PSNR and SSIM measures obtained on the GoPro test split is provided in Table 1. It can be observed from the quantitative measures that our method performs better compared to previous state-of-the-art.

\textbf{Qualitative Evaluation:}
Visual comparisons on different dynamic and 3D scenes are shown in Figs.~\ref{fig:dynamic}.
Visual comparisons  are given in Fig.~\ref{fig:dynamic}. We observe that the results of prior works suffer from incomplete deblurring or artifacts. In contrast, our network is able to restore scene details more faithfully which are noticeable in the regions containing text, edges, etc. An additional advantage over \cite{hyun2013dynamic,whyte2012non} is that our model waives-off the requirement of parameter tuning during test phase. The proposed method achieves consistently better PSNR, SSIM and visual results with lower inference-time than DMPHN (\cite{zhang2019deep}).

\section{Conclusion}
We proposed a new content-adaptive architecture design for the challenging task of removing spatially-varying blur in images of dynamic scenes. Efficient self-attention is utilized in all the encoder-decoder to get better representation whereas cross-attention helps in efficient feature propagation across layers and levels. Proposed dynamic filtering module shows content-awareness for local filtering. The proposed method is more interpretable which is one of its key strengths. Our experimental results demonstrated that the proposed method achieved better results than state-of-the-art methods on two benchmarks both qualitatively and quantitatively. We showed that the proposed content-adaptive approach achieves an optimal balance of memory, time and accuracy and can be applied to other image-processing tasks. Refined and complete version of this work appeared in CVPR 2020.

\bibliographystyle{unsrtnat}
\bibliography{references}  %%% Uncomment this line and comment out the ``thebibliography'' section below to use the external .bib file (using bibtex) .

\end{document}